\documentclass[aps,twocolumn,prl]{revtex4}

\usepackage{graphicx}
\def\<{\langle}
\def\>{\rangle}
\begin{document}
\title{Vibrational edge modes in intrinsically heterogeneous doped transition metal oxides}
\author{I. Martin, E. Kaneshita, and A. R. Bishop}
\affiliation{
Los Alamos National Laboratory, Los Alamos, NM 87545}
\author{R. J. McQueeney}
\affiliation{
Ames Laboratory, Ames, IA 50011}
\author{Z. G. Yu}
\affiliation{SRI International, Menlo Park, CA 94025}
\date{\today}
\begin{abstract}
By applying an unrestricted Hartree-Fock and a Random Phase approximations to a multiband Peierls-Hubbard Hamiltonian, we study the phonon mode structure in models of transition metal oxides in the presence of
intrinsic nanoscale inhomogeneities induced by hole doping.  We identify low
frequency $local$ vibrational modes pinned to the sharp interfaces between
regions of distinct electronic structure (doped and undoped) and separated in frequency from the
band of extended phonons. A characteristic of these ``edge'' modes is
that their energy is essentially insensitive to the doping level.
We discuss the experimental manifestations of these modes in inelastic neutron scattering, and also in spin and charge excitation spectra.

\end{abstract}
\maketitle

Recent advances in experimental techniques provide strong indications that a
broad class of electronic materials, commonly referred to as {\em strongly
correlated}, exhibit intrinsically heterogeneous electronic
phases\cite{1,2,3,4,5,6,6.5,7}. Strong interactions in these materials drive phase
separation between regions of distinct electronic structure\cite{dagotto}.
However, a competing interaction or electronic kinetic energy can lead to
frustration of the global phase separation and to formation of {\em nanoscale
inhomogeneities}, which can take the form of, e.g., stripes/checkerboards\cite{zaanen} or clumps\cite{koulakov}.
Unlike the conventional spin and charge density wave phases, the ordering of intrinsic heterogeneities is sensitive to thermal and quantum fluctuations\cite{fradkin}, as well as static disorder.
This makes an unambiguous identification of such phases, and their distinction from the more familiar density wave instabilities problematic, requiring development of new diagnostics more sensitive to the unique structure of such phases.
Here, we propose one such diagnostic tool which relies on detection of localized {\em edge modes} pinned to the interface between the regions of different electronic
structure.  The edge (interface) modes appear in every degree of freedom
coupled to the inhomogeneity (charge, spin, lattice)\cite{yuPRB}, with corresponding correlated experimental signatures.
Here we study the {\em lattice vibrational edge modes} which lend themselves to study with inelastic neutron scattering.  We also compute corresponding excitation spectra in spin and charge channels and describe the cross-correlations.

We focus on one particular class of transition metal oxides that has attracted significant attention recently, namely cuprates.
These are the materials exhibiting high-$T_c$ superconductivity upon doping.
The presence of stripes/checkerboards and their possible influence on superconductivity are subjects of intensive debate.
Since it appears that stripes can significantly affect
superconductivity\cite{kivel,martin,bishop}, it is important to establish whether they are indeed present in the materials.
Our purpose here is to elucidate typical manifestations of stripe order in experimental probes sensitive to the excitations in spin, lattice and charge sectors.
In particular, we compare our theoretical results with the observations of ``anomalous'' phonon modes in the cuprates\cite{robCu1,robCu2,ybco}.
Analogous results also apply to edge mode signatures in other doped transition metal oxides (and related complex electronic materials), as found in recent experimental studies\cite{Ni,Bi,eCu,CMR}; corresponding modes (phonon shape modes) have already been predicted and identified in conjugated polymers and quasi-1D (one-dimensional) charge transfer solids\cite{Raman,RRaman}.

%%%% formulation
To model the CuO$_2$ planes of doped cuprates, we use a 2D three band extended Peierls-Hubbard Hamiltonian, which includes both electron-electron and electron-phonon interactions\cite{yone,yi}:
\begin{eqnarray}\label{eq:H0}
H_0 = \sum_{<ij> \sigma} t_{pd}(u_{ij})
(c^\dagger_{i\sigma}c_{j\sigma}+ H.c.)
+\sum_{i,\sigma} \epsilon_i(u_{ij})
c^\dagger_{i\sigma}c_{i\sigma} \nonumber\\
+\sum_{<ij>} \frac{1}{2} K_{ij}u^2_{ij}
+\tilde{\sum_{i, j,\sigma, \sigma'}}\frac{1}{2}{U_{ij}n_{i\sigma}n_{j\sigma'}}.
\end{eqnarray}
Here, $c^\dagger_{i\sigma}$ creates a hole with spin $\sigma$ on site $i$;
each site has one orbital (d$_{x^2-y^2}$ on Cu, or O p$_x$ or p$_y$ on O). The
Cu (O) site electronic energy is $\epsilon_d$ ($\epsilon_p$). Here, $U_{ij}$
represents the on-site Cu (O) Coulomb repulsion, $U_d$ ($U_p$), or the
intersite one, $U_{pd}$; and the summation $\tilde{\sum}$ is taken except for the case $(i,\sigma)=(j,\sigma')$. The electron-lattice interaction causes modification
of the Cu-O hopping strength through the oxygen displacement $u_{ij}$:
$t_{pd}(u_{ij}) = t_{pd}\pm \alpha u_{ij}$, where $+(-)$ applies if the Cu-O
bond shrinks (stretches) for a positive $u_{ij}$; it also affects the Cu
on-site energies $\epsilon_d(u_{ij}) = \epsilon_d + \beta\sum_j{(\pm u_{ij})}$,
where the sum runs over the four neighboring O ions. The other oxygen modes
couple to electron charge more weakly, and are neglected for simplicity.
We use the following set of model parameters\cite{yone,yuPRB}:
$\epsilon_p-\epsilon_d = 4.4$ eV, $U_d =11$ eV, $U_p = 3.3$ eV, $U_{pd} = 1.1$ eV, and $K = 38.7$ eV/{\AA$^2$}, $\alpha = 5.2$ eV/\AA, $\beta = 1.2$ eV/\AA, with $t_{pd} = 1.1$ eV.
To approximately solve the model, we use  unrestricted Hartree-Fock (HF) combined with an inhomogeneous generalized Random Phase Approximation (RPA) for linear lattice fluctuations\cite{yone} in a supercell of size $N_x\times N_y$ with periodic boundary conditions.
In this model, doped holes tend to segregate into the stripes due to competition between the magneto-elastic interaction that favors global electronic phase separation and electronic kinetic energy that
favors uniform carrier density.
We note that other competing interactions can produce stripes, clumps and other inhomogeneities\cite{kiv,branko}; the local, coupled charge-spin-lattice dynamics governing edge modes on these templates then can still be modelled with a similar Hamiltonian and RPA analysis.

The output of the calculation here is the inhomogeneous HF groundstate and the phonon eigenfrequencies and eigenvectors.
From the phonon eigenmodes, we calculate the corresponding neutron scattering cross section
\begin{eqnarray}
S(\mathbf{k},\omega) =\int dt\, e^{-i\omega t}
\sum_{l l^\prime}{\langle e^{-i \mathbf{k R}_l(0)}
e^{i\mathbf{k R}_{l'}(t)}\rangle},
\end{eqnarray}
where $\mathbf{R}_{l}(t) =  \mathbf{R}_{l}^0 + \mathbf{d}_l + \mathbf{u}_l(t)$
is the position of the $l$-th oxygen atom expressed in terms of the location of
the unit cell origin $\mathbf{R}_{l}^0$, position within the unit cell
$\mathbf{d}_l$, and time-dependent vibrational component $\mathbf{u}_l(t)$. For
phonon modes with $\mathbf{u}_l(t)$ oriented along the corresponding
metal-oxygen bonds, on the O$_x$ sublattice $\mathbf{d}_l = \frac a2 \hat{x}$
and $\mathbf{u}_l\equiv x_l \hat{x}$, and on the O$_y$ sublattice $\mathbf{d}_l
= \frac a2 \hat{y}$ and $\mathbf{u}_l\equiv y_l \hat{y}$. The scalar
displacements can now be expressed in terms of the normal modes $z_n$ as
$x_l(t) = \sum_n{\alpha_{x_l,n}z_n(t)}$ and $y_l(t) =
\sum_n{\alpha_{y_l,n}z_n(t)}$. Making first order expansion in the oxygen
displacements, we obtain
\begin{eqnarray}
S(\mathbf{k},\omega) &=& \sum_n
\Big[k_x^2|\alpha_{\mathbf{k},n}^x|^2
+ k_y^2|\alpha_{\mathbf{k},n}^y|^2\\\nonumber
&& + k_x k_y(e^{i(k_x-k_y)a/2}
\alpha_{\mathbf{k},n}^x
\alpha_{-\mathbf{k},n}^y + c.c.)\Big]\\\nonumber
&&\times\frac{\hbar}{2m\omega_n}
[(1+n_B)\delta(\omega-\omega_n) + n_B\delta(\omega+\omega_n)].
\end{eqnarray}
Here, $\alpha_{\mathbf{k},n}^x  = \sum_l{e^{-i\mathbf{k
R}_l^0}\alpha_{x_l,n}}$, and $n_B = (e^{\omega_n/T}-1)^{-1}$ is the thermal
population of the phonon mode $n$. This is a generalization of the usual
neutron scattering intensity expression\cite{lovesay} for the case of phonons
with a larger real space unit cell.  We plot
$S(\mathbf{k},\omega)/|\mathbf{k}|^2$ for \textbf{k}-directions sampling
longitudinal modes, consistent with the common experimental convention.

\begin{figure}[htbp]
\begin{center}
\includegraphics[width = 0.9\linewidth]{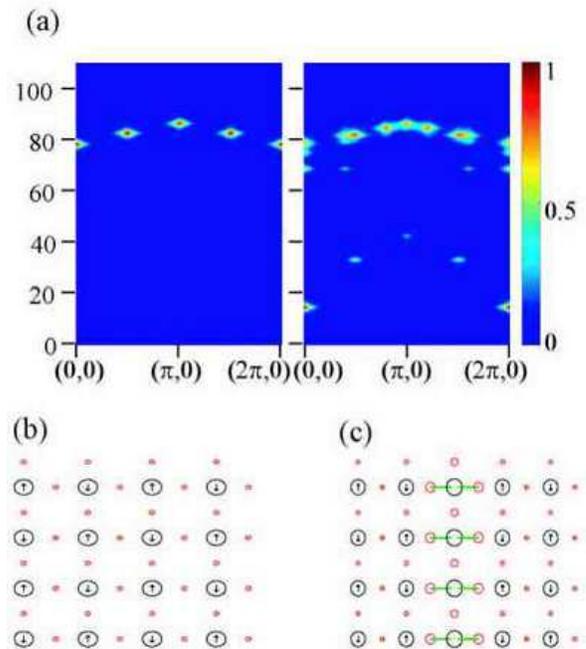}
\caption{(a) Neutron scattering  spectra and the ground state structures for the undoped (left panel in (a), and (b)), and hole-doped (20\%) case (right panel in (a), and (c)).
The neutron scattering spectra are symmetrized with respect to directions parallel and perpendicular to the stripe.
In the lower plots, black circles represent Cu sites and red circles represent O.
The radius of the circle is proportional to the corresponding site hole density.
Arrows centered on circles show the magnitude and direction of spin.  The green lines originating from O sites indicate the magnitude and direction of the equilibrium O displacements in the presence of a stripe.} \label{fig:Cu}
\end{center}
\end{figure}

%%% results 1
We first analyze the signatures  of the edge mode by comparing computed phonon spectra for doped and undoped cuprates (Fig.~\ref{fig:Cu}).
The predicted neutron scattering intensities are given in the original Brillouin zone, symmetrized with respect to the two possible orientations of the stripe.

\begin{figure}[htbp]
\begin{center}
\includegraphics[width = 0.8 \linewidth]{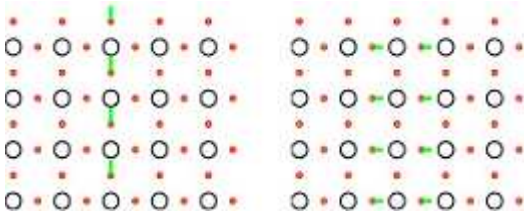}
\caption{Localized vibrational eigenmodes for the case of a 20 \%
 doped CuO$_2$ plane.
The stripe is centered along the middle vertical Cu row.
There are two localized branches:  One corresponds to the oxygen vibration parallel to the stripe (low frequency, $E = 14.8$ meV, left), and one corresponds to the oxygen displacements perpendicular to the stripe (high frequency, $E = 68.5$ meV, right). }
\label{fig:Cu_mode}
\end{center}
\end{figure}

The ground state configuration in the undoped case is a uniform antiferromagnet (AF), while, in the doped (20\%)  case, the holes partially segregate into a Cu-centered vertical stripe \cite{foot1}.
The doping produces $two$ low-frequency branches that split off from the band of extended phonon modes
characterizing the undoped state.
In Fig.~\ref{fig:Cu_mode} we plot representative modes from these branches.
Clearly, these are modes localized around the stripe.
The dispersive lower frequency band centered around $\sim 30$ meV corresponds to the oxygen displacements \textit{along} the stripe and the narrow high frequency one at $\sim 70$ meV to the vibrations \textit{perpendicular} to the stripe.
(The example shown in Fig.~\ref{fig:Cu_mode}, right, is a localized
asymmetric breathing mode\cite{robCu1,robCu2}.)
Experimental evidence from neutron scattering\cite{robCu1,robCu2} shows mode softening of about $15$ percent with doping in the cuprates, consistent with the softening obtained here for the $perpendicular$ edge mode.
Recent neutron measurements in twinned and detwinned YBa$_2$Cu$_3$O$_{6+\delta}$ \cite{ybco} have attempted to resolve the phonon spectrum both parallel and perpendicular to the purported stripe direction\cite{mook}, but conclusive identification has been hampered by the degeneracy and mixing of many modes near the zone boundary.
There has been no search yet for the $parallel$ edge modes which are predicted here to have a huge softening of ${\sim 60}$ percent.
However, there are distinct anomalies in the longitudinal phonon spectrum of La$_{2-x}$Sr$_x$CuO$_4$ around 30 meV that have not been studied in detail\cite{robCu2}.

Despite the close agreement between the energy scales of the predicted and observed modes, it is important to note some deviations from the reported experimental data for the mode intensities.
We speculate that these discrepancies may be due to the detailed local oxygen environments:  Either (a) more complex intra-unit cell distortions, i.e. local
oxygen polarizations (e.g., from local confinement or pairing effects), which
would require more detailed modelling than the constrained O displacements used
here, or (b) more complex stripe configurations than the regular linear stripes
we have considered. (Such textured, e.g., checkerboard, patterns are indeed found in
certain models\cite{dagotto,branko}).

%%% results 2
We now investigate the structure of corresponding electronic excitations in the \textit{charge} and \textit{spin} channels.
We again use the inhomogeneous generalized RPA; however, due to the increased computational complexity, we must use smaller system sizes:
A $3\times3$ system for diagonal stripes and a $3\times4$ system for vertical stripes.
In both cases the hole concentration is $\frac{1}{3}$.
We have analyzed the spectra, as well as the real-space structure of these excitations.
For example, the excitation spectrum in the longitudinal spin channel is given by $I(\omega,q)=\sum_n{|\<0| S_z(q) |n\>|^2\delta(\omega-E_n+E_0)}$, where $|0\>$ and $|n\>$ are the Hartree-Fock ground state and an RPA excited state, respectively.
The real-space structure of the excitation $|n\>$ is described by, e.g., $\<0| S_z(i) |n\>$.
This quantity shows how the expectation value of the $z$-component of spin on site $i$ changes due to to the excitation of the RPA state $|n\>$ when the system is driven by an external field of frequency $E_n-E_0$.

The phonon spectrum for the vertical $3\times4$ case is essentially the same as in the $4\times5$ case in Fig.~\ref{fig:Cu}, with two localized branches  near 30 meV and 70 meV.
This insensitivity to doping is a characteristic of the edge modes.
For the $3\times3$ diagonal stripe  case, the phonon spectrum has only the high-energy edge mode at 70 meV, but no excitation near 30 meV.
Note that the low-energy vibrational mode can only exist for a vertical stripe because the oxygen-mediated charge transfer along the stripe is suppressed in the diagonal case.

\begin{figure}[htbp]
\begin{center}
\includegraphics[width=0.9\linewidth]{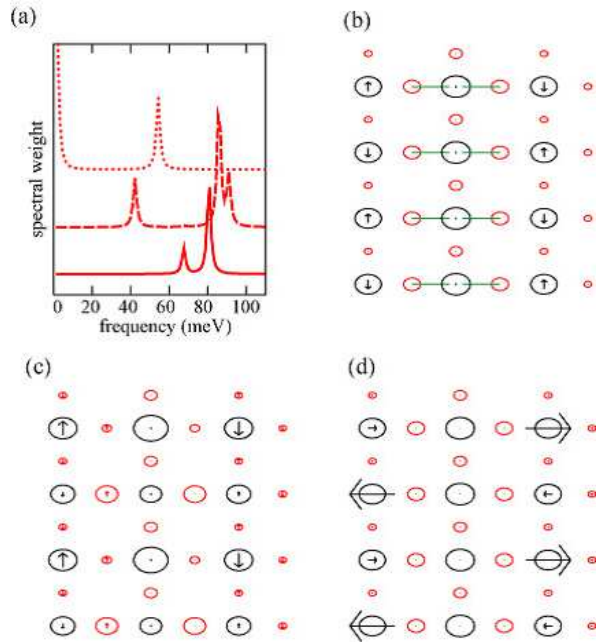}
\caption{$3\times4$ system with a vertical stripe. (a) Spectral weights of longitudinal magnetic excitation (solid lines), transverse magnetic excitation (dotted line) and charge excitation (dashed lines); blue lines are for (0,0) and red lines for $(\frac{2}{3}\pi,\pi)$ momentum transfers.
The scale for the transverse spin mode is 4,000 times larger than the one for others. The magnified spectrum for the charge excitation at $(\frac{2}{3}\pi,\pi)$ is shown in the right panel.
(b) The ground state. (c) Spin and charge configurations for the excited state $\sim$ 40 meV; this frequency corresponds to the oxygen vibration parallel to the stripe.  Here $S_x=0$.
(d) $S_x$ configuration for the excited state $\sim$ 50 meV, where $S_z=0$} \label{fig:Sqw-v}
\end{center}
\end{figure}

The  spectra for magnetic and charge excitations for the vertical stripe case are presented in Fig.~\ref{fig:Sqw-v}.  The spectral weight of the charge excitation at $q= (\frac{2}{3}\pi,\pi)$ has a peak near $\sim$ 40 meV, which corresponds to the frequency of the ``half-breathing" oxygen vibration along the stripe (same low-energy branch as Fig.~\ref{fig:Cu_mode}a, but with $q_y = \pi$). This charge excitation produces dynamic dimerization along the stripe,
which couples to the anti-phase oxygen vibration parallel to the stripe. On the other hand, the excitation near 70 meV is associated with the anti-phase oxygen vibration perpendicular to the stripe (same split-off branch as Fig.~\ref{fig:Cu_mode}b), and corresponds to the charge excitation at the Cu sites next to the stripe.
There is also a magnetic longitudinal excitation at momentum $(0,0)$ near 70 meV, which is coupled to the anti-phase oxygen vibration perpendicular to the stripe.
The transverse spin channel mode at $(\frac{2}{3}\pi,\pi)$ shows not only the Goldstone zero-mode but also a higher energy excitation $\sim 50$ meV.
From Fig.~\ref{fig:Sqw-v} (d), this higher energy excitation likely corresponds to the spin wave branch folded back at the zone boundary (larger systems sizes are necessary to be definitive.)
From continuity of spin wave excitations, it follows that there is an intensity somewhere below 50 meV at $(\pi,\pi)$.
It is possible that this mode has a relationship with the ``41 meV" magnetic resonance mode\cite{rosat,batista}.
Note that there is no coupling of magnetic excitations with the low-energy vibrational mode.
This is because in the Cu-site centered stripe oxygen vibration only transfers charge between the sites with no spin polarization, and hence does not couple to the spin channel.

\begin{figure}[htbp]
\begin{center}
\includegraphics[width = 0.9 \linewidth]{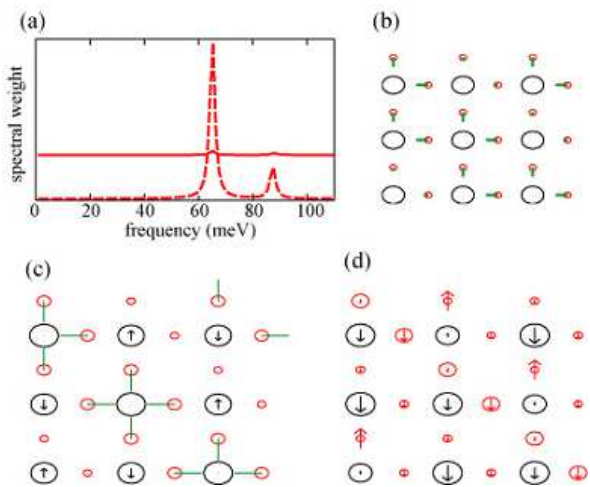}
\caption{$3\times3$ system with diagonal stripe. (a) Spectral weights of longitudinal magnetic excitations (solid lines) and charge excitations (dashed lines); blue lines are for the spectra at momentum (0,0) and red lines for the ones at $(\frac{2}{3}\pi,\frac{2}{3}\pi)$. (b) The phonon mode at 64.3 meV. (c) The ground state and (d) the excited state $\sim 65$ meV.} \label{fig:Sqw-d}
\end{center}
\end{figure}

Fig.~\ref{fig:Sqw-d} (a) shows the spectral weights for excitations in the diagonal stripe case.
Here, there are several RPA excited states near 65 meV, which
lead to a peak in the longitudinal spin excitation spectrum.  The excited state
that has the lowest energy of the states $\sim$ 65 meV is plotted in
Fig.~\ref{fig:Sqw-d} (d); the vibrational mode corresponding to this excitation
is shown in Fig.~\ref{fig:Sqw-d} (b).
As discussed earlier, unlike for the vertical stripe case there
is no localized vibrational mode around 40 meV, and accordingly there are no RPA spin or charge excitations in this energy range.
The only excitations of the longitudinal magnetic and charge channels appear in the 65 $\sim$ 90 meV range.
The transverse magnetic excitation shows only the Goldstone mode.

%%% conclusion
We conclude that the formation of split-off $local$ modes in spin, charge and lattice channels, familiar in the polaronic physics of conjugated polymers and quasi-1D charge transfer salts\cite{poly,Raman,RRaman}, is also expected to be prominently manifested in the case of other doped strongly correlated materials, including nickelates, cuprates, and bismuthates\cite{1,2,3,4,5,6,6.5,7}.
The intensities of these signatures should increase with doping (and therefore density of interfaces).
In addition to ARPES and inelastic magnetic and lattice neutron scattering, experimental techniques that should be applied to study the edge modes include dielectric constants, and infra-red and Raman spectroscopy\cite{Raman}; resonant Raman scattering would be particularly valuable even for low doping levels\cite{RRaman}.

We would like to acknowledge valuable discussions with T. Egami.  This work was supported by the U.S. DOE.


\begin{thebibliography}{99}
\bibitem{foot1} In the present model the most stable stripe configuations corresponds to the insulating stripes with linear filling 1, while experimentally observed stripe periodicity may be more consistent with metallic stripes with filling of approximately 1/2\cite{6}.
However, for the edge mode properties, we believe, this distinction is not crucial.
\end{thebibliography}
\end{document}